\pdfoutput=1

\documentclass[11pt]{article}

\usepackage[preprint]{acl}

\usepackage{times}
\usepackage{latexsym}

\usepackage[T1]{fontenc}

\usepackage[utf8]{inputenc}

\usepackage{microtype}

\usepackage{inconsolata}

\usepackage{graphicx}
\usepackage{amsmath}
\usepackage{amssymb}
\usepackage{makecell}
\usepackage{multirow}
\usepackage{threeparttable}
\usepackage{pifont}
\usepackage{ulem}
\usepackage{subcaption}

\title{LangGPT: Rethinking Structured Reusable Prompt Design Framework for LLMs from the Programming Language}

\author{
 \textbf{Ming Wang\textsuperscript{1}},
 \textbf{Yuanzhong Liu\textsuperscript{2}},
 \textbf{Xiaoyu Liang\textsuperscript{3}},
 \textbf{Songlian Li\textsuperscript{2}},
 \textbf{Yijie Huang\textsuperscript{1}}, \\
 \textbf{Xiaoming Zhang\textsuperscript{1}},
 \textbf{Sijia Shen\textsuperscript{4}},
 \textbf{Chaofeng Guan\textsuperscript{4}},
 \textbf{Daling Wang\textsuperscript{1}},
 \textbf{Shi Feng\textsuperscript{1}}, \\
 \textbf{Huaiwen Zhang\textsuperscript{1}},
 \textbf{Yifei Zhang\textsuperscript{1}},
 \textbf{Minghui Zheng\textsuperscript{3}},
 \textbf{Chi Zhang\textsuperscript{1}},
\\
\\
 \textsuperscript{1}Northeastern University, 
 \textsuperscript{2}Wuhan University,
 \textsuperscript{3}Zhejiang University,
 \textsuperscript{4}Shenyang Aerospace University
\\
 \small{
   \textbf{Correspondence:} \href{mailto:wangdaling@cse.neu.edu.cn}{wangdaling@cse.neu.edu.cn},
   \href{mailto:liu.yuanzhong@foxmail.com}{liu.yuanzhong@foxmail.com}
 }
}

\begin{document}
\maketitle
\begin{abstract}
LLMs have demonstrated commendable performance across diverse domains. Nevertheless, formulating high-quality prompts to instruct LLMs proficiently poses a challenge for non-AI experts. Existing research in prompt engineering suggests somewhat scattered optimization principles and designs empirically dependent prompt optimizers. Unfortunately, these endeavors lack a structured design template, incurring high learning costs and resulting in low reusability. In addition, it is not conducive to the iterative updating of prompts. Inspired by structured reusable programming languages, we propose LangGPT, a dual-layer prompt design framework as the programming language for LLMs. LangGPT has an easy-to-learn normative structure and provides an extended structure for migration and reuse. Experiments illustrate that LangGPT significantly enhances the performance of LLMs. Moreover, the case study shows that LangGPT leads LLMs to generate higher-quality responses. Furthermore, we analyzed the ease of use and reusability of LangGPT through a user survey in our online community\footnote{The homepage: \href{http://www.langgpt.ai}{https://langgpt.ai}, the Github project: \href{https://github.com/langgptai/LangGPT}{https://github.com/langgptai/LangGPT}, the tools and experiments: \href{https://github.com/sci-m-wang/LangGPT-tools}{https://github.com/sci-m-wang/LangGPT-tools}, and the Huggingface organization: \href{https://huggingface.co/langgptai}{https://huggingface.co/langgptai}}.
\end{abstract}

\section{Introduction}

Large language models (LLMs) can execute diverse tasks \citep{sun_survey_2023,sun_evaluating_2023,yu_large_2023} based on powerful language comprehension, reasoning, and generation capabilities. Injecting domain knowledge also enables LLMs to perform domain-related specific tasks \citep{wang2023huatuo,li2023chatdoctor}. Thus, LLM-based applications have attracted the attention of many non-AI experts, who have shown keen interest in using LLMs for programming, writing articles, medical diagnostics, etc \citep{10.1145/3544548.3581388}. However, fully unleashing these capabilities of LLMs requires premium quality prompts \citep{eric_complete_2022,chen_unleashing_2023,nagaraju_gajula_guide_2023}, but non-AI experts, even some AI-related practitioners, may not have the experience to write such prompts. Therefore, prompt engineering has been widely focused on \citep{tanay_varshney_introduction_2023,mesko_prompt_2023,zhenxuan_wang_how_2023}.

Prompt engineering is a typically empirical science, which mainly involves designing and optimizing the prompts. As LLMs can understand natural language, it is possible to ask them to execute tasks through unstructured natural language instructions directly. On this basis, some tricks for prompt optimization were first explored by researchers. \citet{bsharat_principled_2023} introduce 26 guiding principles designed to improve the performance of LLMs. \citet{liu_design_2022} have evaluated 5493 generations covering 51 themes and 51 styles throughout five experiments in the text-to-image task and summaries prompt design guidelines. In addition to these tricks, some researchers have also focused on optimizing prompts based on historical data. \citet{sun_autohint_2023} guide LLMs to derive new prompts for a given instance from the incorrect reasoning, and then summarise the corresponding prompts for each instance as a reference for optimizing prompts. \citet{pryzant_automatic_2023} leverage mini-batches of data to form natural language ``gradients'' and utilize beam search and bandit selection procedure to edit the current prompt in the opposite semantic direction of the gradient. \citet{fan_towards_2023} analyze a large prompt database and present an automatic prompt optimization framework.

Direct prompt optimization principles and methods based on historical data require a wealth of experience. Moreover, these usually perform well only for specific tasks or domains. To improve generalization, some researchers have proposed adaptive prompt optimization methods. \citet{guo_connecting_2023} connect LLMs with evolutionary algorithms and proposes a novel framework for discrete prompt optimization, called EvoPrompt. \citet{li_dialogue_2023} design a multi-round dialogue alignment strategy and utilizes GPT-4 \citep{achiam_gpt-4_2023} to generate a readability prompt set. Meanwhile, they propose an efficient prompt screening metric that can filtrate high-quality prompts with linear complexity. \citet{wang_promptagent_2023} introduce PromptAgent which can reflect on model errors and generate constructive error feedback to induce precise expert-level insights and in-depth instructions. \citet{hao_optimizing_2022} and \citet{cheng_black-box_2023} optimize prompts by aligning human and LLMs' preference styles.

Prompt optimization can significantly improve the performance of LLMs. However, such fragmented, unstructured prompt optimization principles lead to low reusability of quality prompts. In addition, it is not conducive to iterative updating of prompts. Consequently, some researchers have devised rules for the construction of prompts. \citet{nigh_chatgpt3-free-prompt-list_nodate} collects vast quality prompts and summaries of the CRISPE rule for prompt design. \citet{zamfirescu-pereira_why_2023} take a prototype LLM-based chatbot design tool as the design probe, supporting non-AI-experts engage in ``end-user prompt engineering''. Some researchers have designed prompt construction rules for applications in different domains. \citet{cao_study_2023} present various prompt templates on deep learning program repair tasks for ChatGPT. \citet{yeh_decorate_2022} reformulate the biomedical relation extraction task as a cloze-test task under a simple prompt formulation.

These methods, which are based on a great deal of experience in use, have strong domain relevance and model relevance, with low generalisability, flexibility, and reusability. To further unleash the performance of LLMs, some researchers have defined agents. Agents empower LLMs to use tools, acquire domain knowledge, retain long-term or short-term memories, and plan \citep{xu_exploring_2023,xi_rise_2023,park_generative_2023}. Although agent methods \citep{hong_metagpt_2023,wu_autogen_2023} have systematically designed the key components of the prompt and reserved flexible customization interfaces, the learning costs are very high. In addition, it is difficult for non-AI experts to modify agent designs.

\begin{table*}[h]
\begin{tabular}{c|m{0.4\textwidth}m{0.4\textwidth}}
\hline
Aspects                 & Natural Language                                                                                                                                                                                                                                               & Programming Language                                                                                                                                                                                                           \\ \hline
Audience        & Natural languages are spoken by humans to humans \citep{grune_modern_2012}.                                                                                                                                                                                        & Programming languages are prepared by humans for machines \citep{chakray_programming_2018}.                                                                                                                  \\ \hline
Structure              & Natural languages have loose and flexible syntax and semantics, permitting creativity and variation and possessing a high degree of fault tolerance.                                                                                                                                 & Computers can only understand fixed instructions and require programming languages to have strict, rigorous syntax and semantics.                                                                                                         \\ \hline
Ambiguity              & Natural languages are more ambiguous, but humans can clarify the meaning of expressions \citep{chakray_programming_2018,aho_compilers_2007}.                                                                                                     & Programming languages are less ambiguous because they need to provide computers with clear instructions \citep{geeksforgeeks_natural_2023}.                                                                                       \\ \hline
Evolution & Natural languages evolve naturally over time through human use and communication \citep{sipser_introduction_1996}. Natural languages are flexible in adding new words and meanings and discarding outdated usages \citep{fromkin2018introduction}. & Programming languages are specifically designed to communicate with machines \citep{sebesta_concepts_2012}. New syntax rules and features require explicit upgrades or releases \citep{pratt_programming_1984}. \\ \hline
\end{tabular}
\caption{Difference between programming languages and natural languages}
\label{tab:diff}
\end{table*}

To promote LLM-based applications and further stimulate the potential of LLMs, we would like to design a generalizable, reusable, and extensible prompt template. In addition, the template ought to be easy to learn and easy to work with for non-AI experts. Inspired by the belief that prompt is the programming language of the LLM era \citep{alouani_prompt_2023,mund_ai_2023}, we have designed \textbf{Lang}uage for \textbf{GPT}-like LLMs (LangGPT), a prompt design framework. LangGPT refers to the systematic, prescriptive, and reusable properties of programming languages, and retains the flexibility and extensibility of natural languages. We analyzed the differences between natural languages and programming languages and designed LangGPT as a dual-layer structure, which consists of modules and internal elements. Modules in LangGPT can be divided into two categories: inherent modules and extension modules. For inherent modules, we design the necessary internal elements of each module in detail and give example templates. For the extension modules, we unified the design of the basic internal elements. Experiments and case studies have demonstrated that LangGPT is better than baseline prompts for bootstrapping LLMs. Furthermore, with the belief that better instruction understanding leads to greater performance of LLMs \citep{ouyang_training_2022,chung_scaling_2024}, we constructed a dataset for LangGPT prompt generation. Experiments demonstrate that LLMs trained on the LangGPT prompt tasks have significantly improved performance. In addition, we conducted user research to prove the ease of use and user satisfaction of LangGPT based on our community.

In summary, the contributions of this work include: (1) We proposed a dual-layer structured prompt design framework LangGPT to improve the generalization and reusability of prompts, reducing the learning cost of prompt design. (2) We demonstrated through experiments and case studies that LangGPT prompts can better guide LLMs in executing tasks. Meanwhile, we experimentally proved that training on generating LangGPT prompts can improve the performance of LLMs. (3) We have built an online community based on LangGPT and conducted a user survey to verify the ease of use and reusability of LangGPT.

\section{Prompt Design Principles with Programming Language}
Programming languages are more standardized and reusable compared to natural languages. To design high-quality reusable prompts, we analyzed the differences between natural languages and programming languages and proposed prompt design principles.
\subsection{Difference between Programming Languages and Natural Languages}
While natural languages are primarily used for communication, programming languages are designed to instruct machines to execute tasks \citep{geeksforgeeks_natural_2023}. The different purposes have led to very different contexts for the creation and evolution of the two languages. The main differences between these languages are shown in Table \ref{tab:diff}.


\begin{figure*}
    \centering
    \includegraphics[width=0.9\textwidth]{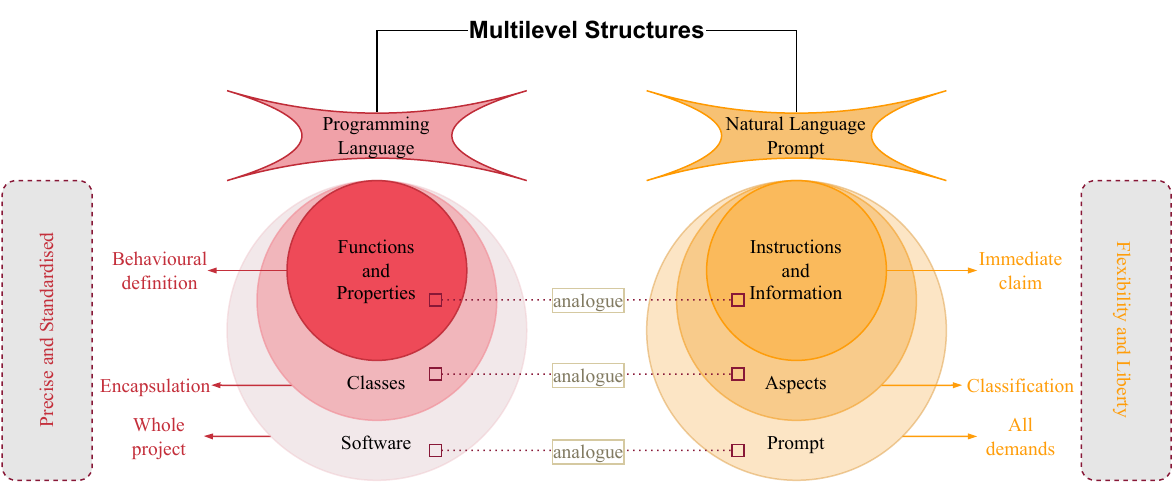}
    \caption{Analogy between programming language and natural language prompt. The analogy between the two types of languages was analyzed in terms of their hierarchical structure. Circles of different sizes indicate different layers. Smaller circles indicate closer to the inner layers, corresponding to darker colors.}
    \label{fig:correspondence}
\end{figure*}

In summary, the main difference between the two languages is that natural languages are more vague and flexible, while programming languages are more standardized and precise. LLMs essentially perform a large number of computations and share many similarities with machines. Therefore, we propose LangGPT, a natural language programming framework for LLMs, by drawing on the characteristics of programming languages and combining the advantages of natural languages.

\subsection{Principles for Prompt Design}
We improve prompts by referring to the design ideas of programming languages. After analyzing the differences between programming languages and natural languages, we propose the design principles for prompts: (1) \textbf{Prompts should have a regularised format}. Flexible and ambiguous natural languages are difficult to understand for LLMs. Format-constrained prompts make users' purpose and requirements more salient. (2) \textbf{The structure of prompts should be extensible}. Custom structures facilitate users to design suitable prompts according to their domains and tasks. (3) \textbf{Specific requirements must be clear and complete}. Both instructions and additional requirements should be explicit and complete to avoid misunderstandings or biases. (4) \textbf{Languages should be flexible}. Where the requirements are clear, flexible languages can be better adapted to different domains. In addition, flexible languages are easy for users to learn.

\begin{table*}[h]
\renewcommand\arraystretch{1.2}
\centering
\resizebox{\textwidth}{!}{
\begin{tabular}{lccccccccccc}
\hline
\textbf{Scenario}               & \textit{Prof.}             & \textit{Cons.}               & \textit{Goal}                     & \textit{Init.}            & \textit{Ex.}                  & \textit{Wkflo.}                  & \textit{Skill}                     & \textit{Sug.}         & \textit{Bkgrd.}          & \textit{Style}                      & \textit{Outf.}               \\
\hline
Writing               & \checkmark & \checkmark & \checkmark & \checkmark & \checkmark & \checkmark  & \checkmark  & \ding{55} & \checkmark  & \checkmark  & \checkmark  \\
Role-playing           & \checkmark & \checkmark & \checkmark & \checkmark & \checkmark & \ding{55} & \checkmark  & \checkmark  & \checkmark  & \checkmark  & \ding{55} \\
Entertainment          & \checkmark & \checkmark & \checkmark & \checkmark & \checkmark & \ding{55} & \checkmark  & \checkmark  & \ding{55} & \checkmark  & \ding{55} \\
Supplementary Learning & \checkmark & \checkmark & \checkmark & \checkmark & \checkmark & \checkmark  & \checkmark  & \checkmark  & \checkmark  & \checkmark  & \checkmark  \\
Prompt Optimisation    & \checkmark & \checkmark & \checkmark & \checkmark & \checkmark & \checkmark  & \ding{55} & \checkmark  & \ding{55} & \ding{55} & \checkmark  \\
Prompt Hacking         & \checkmark & \checkmark & \checkmark & \checkmark & \checkmark & \checkmark  & \ding{55} & \checkmark  & \ding{55} & \ding{55} & \checkmark  \\
Drawing                & \checkmark & \checkmark & \checkmark & \checkmark & \checkmark & \checkmark  & \ding{55} & \ding{55} & \checkmark  & \checkmark  & \ding{55} \\
Business Operation     & \checkmark & \checkmark & \checkmark & \checkmark & \checkmark & \checkmark  & \checkmark  & \checkmark  & \checkmark  & \ding{55} & \checkmark \\
\hline
\end{tabular}
}
\caption{Status of inherent module definitions. The table lists the eight categories of application scenarios we have defined so far and the modules defined for these scenarios. A \checkmark indicates that a corresponding module has been designed for this scenario. In contrast, a \ding{55} indicates that it was not designed.}
\label{tab:module_definiton}
\end{table*}

\section{Natural Language Programming Framework of Prompts for LLMs}
Based on the design principles, we present LangGPT, the natural language programming framework with a dual-layer architecture for LLMs.
\subsection{Overall Dual-layer Structure}
To systematically design prompts that meet the principles, we have made full reference to the design ideas and structures of object-oriented programming languages \cite{rentsch_object_1982,lutz_programming_2010}. We consider the prompt as a software project and analogize the prompt design process with the software development process. The correspondence is shown in Figure \ref{fig:correspondence}.

Based on analogical analyses, it can be found that natural language prompts have a similar multi-level structure as programming languages. Therefore, we refer to the structure of programming languages propose a dual-layer structure for prompt design, and define the notion of module and element for prompts.

A complete prompt contains several modules. Modules are similar to classes in programming languages, and each module represents an aspect of the requirements for LLMs. For instance, prompts can be augmented in terms of constraints, goals, profiles, etc. Within a module, a number of internal elements are included. Elements are similar to functions and properties in programming languages and represent the content of direct and specific instructions to LLMs. For example, ``Output should be no more than 500 words'' could be an element in a prompt that belongs to the module \textbf{constraint}.

A dual-layer structure can be a good way to standardize the formatting of prompts. However, the flexibility of natural languages would be lost if prompts are too strictly required to follow predefined standard modules and internal elements. In addition, it will reduce the generalisability of LangGPT to different tasks in different domains, which is not conducive to the reuse of quality prompts. To solve these problems, we divided the types of modules and elements into prompts. We defined \textbf{inherent module} and \textbf{basic element} as the predefined dual-layer prompt template. Moreover, we constructed \textbf{extension module} and \textbf{custom element} that support customization. We provide both Markdown \cite{gruber_markdown_2012} and JSON \cite{pezoa_foundations_2016} formats for inherent modules and extension modules. Furthermore, we have written basic elements for different modules and defined principles for writing custom elements.

\begin{table*}[]
    \centering
    \begin{tabular}{c|l}
    \hline
    \textbf{Module}                    & \textbf{Samples of Basic Elements}                                               \\ \hline
    Profile                   & $\bullet$ You are a magazine editor.                                     \\ \hline
    Goal                      & $\bullet$ You need to generate a title for the article.                           \\ \hline
    Constraint                & $\bullet$ The length of the title should not exceed 20 words.                     \\ \hline
    \multirow{6}{*}{Workflow} & \#\#\# Extracting the kernel content \\
                              & $\bullet$ For the given article $\langle \textsc{article} \rangle$, please execute the following actions:         \\
                              & \quad $\circ$ Analyse the theme of the article;                                       \\
                              & \quad $\circ$ Detecting the main objects and related things described in the article; \\
                              & \quad $\circ$ Summarising the core content from the article;                          \\
                              & \quad $\circ$ Save the kernel content.                                                \\ \hline
    Style                     & $\bullet$ The style of the title should be formal.                                \\ \hline
    \end{tabular}
    \caption{Examples of basic internal elements of inherent modules in the writing scenario. This prompt leads LLMs to generate a title for a given article. We have chosen five modules as examples i.e. \textbf{profile}, \textbf{goal}, \textbf{constraint}, \textbf{workflow}, and \textbf{style}. Particularly, for the workflow module, we show a function-like element.}
    \label{tab:basic_elements}
\end{table*}

\subsection{Tectonics of Inherent Modules}
The module serves as a connection between the complete prompt and the instruction unit and plays a very important role in controlling the structure of the prompt. 

We define inherent modules for critical aspects that are required for almost all prompts. Furthermore, we define inherent modules for certain scenarios that are relevant to the application for ease of learning and use. Table \ref{tab:module_definiton} shows the inherent modules we defined for some scenarios. 

In Table \ref{tab:module_definiton}, \textit{Prof.} indicates what is expected of LLMs in terms of roles, including \textbf{profile}s, character portraits, etc. \textit{Cons.} denotes \textbf{constraint}s or attention, i.e., ranges that LLMs are not allowed to exceed and requirements that must be met when generating responses, etc. \textit{Goal} lists the \textbf{goal}s that the user wants to achieve, which is what the LLMs need to accomplish. \textit{Init.} is called \textbf{initialization} to inform LLMs that they are about to start a dialogue. Sometimes a specified first sentence is also given in this module. \textit{Ex.} gives LLMs input-output pairs as \textbf{example}s to learn from. \textit{Wkflo.} instructs the \textbf{workflow} when executing a task, similar to the CoT approach \cite{wei_chain--thought_2023}. It is often necessary to instantiate this module when the task requirements are more complex. \textit{Skill} is used to suggest to LLMs the \textbf{skill}s they possess. LLMs that have undergone tool learning can be guided to invoke tools to execute tasks more accurately. In addition, we plan to provide the ability to use tools under this module in future work, with reference to the design of agent tools \cite{chase_langchain_nodate,hong_metagpt_2023}. \textit{Sug.} includes \textbf{suggestion}s and behavioral planning for LLMs. This module focuses on listing common scenarios and giving behaviors or responses that LLMs can take in such situations. \textit{Bkgrd.} indicates the \textbf{background} information and the memories that LLMs are required to have when performing their tasks. \textit{Style} qualifies the \textbf{style} of responses generated by LLMs. \textit{Outf.} defines the  LLMs' \textbf{output-format}. Specifying the output format improves the efficiency and accuracy of the results extraction in certain tasks.

The bolded font is used as the name of the modules in the introduction list.
\subsection{Internal Basic Elements}
Three purposes are typically included in prompts: (1) Implying a certain message to LLMs; (2) Letting LLMs execute a certain task with or without output; (3) The combination of the first two.

The first of these is very similar to the assignment of properties or variables in programming languages. Correspondingly, the last two categories are similar to functions in programming languages. Thus, we construct these three types of basic elements. We use ``The $\langle \textsc{property} \rangle$ is $\langle \textsc{value} \rangle$.'' to simulate an assignment. For the latter two cases, it is necessary to specify the input information, the task, and the output, where input and output can be omitted. We simulate functions using a form like this: ``For the given $\langle \textsc{property} \rangle$ of $\langle \textsc{value} \rangle$, please execute the following actions: $\langle \textsc{actions} \rangle$; Return the $\langle \textsc{result} \rangle$.'' In the basic element writing patterns we provide, the contents contained in the \textbf{angle brackets} need to be populated according to the module and the usage scenario. It is important to note that the writing patterns we have provided only specify the idea of writing internal elements. To improve the generalisability and flexibility of prompts, the language can be adapted to express key information. In Table \ref{tab:basic_elements} we show some examples of the basic elements in some modules.


\subsection{Extension Module and Custom Element}
The inherent modules we have defined have been as comprehensive as possible to cover the many aspects of prompts. Furthermore, we add application scenarios and modules covered by LangGPT. However, limited by our own capabilities and domain knowledge, we were unable to consider all application scenarios of LLMs.

Therefore, in addition to inherent modules and basic elements, we define templates of extension modules and custom elements to improve the generalization and reusability of prompts. 
For application scenarios where inherent modules cannot cover all aspects, new extension modules can be defined as needed. After defining the extension module, the internal elements should also be designed according to the requirements of the extension module. In addition, if inherent modules are compatible with the needs of the scenario but the basic elements cannot meet all of the requirements, custom elements can be added directly to the inherent modules.



\section{Experiments}
To validate the advancement of the proposed LangGPT, we conducted experiments in two aspects: (1) leveraging LangGPT prompts to bootstrap LLMs to perform tasks; and (2) training LLMs on the task of generating LangGPT prompts.

\subsection{Experiment Settings}



\begin{figure}[!h]
    \centering
    \includegraphics[width=0.48\textwidth]{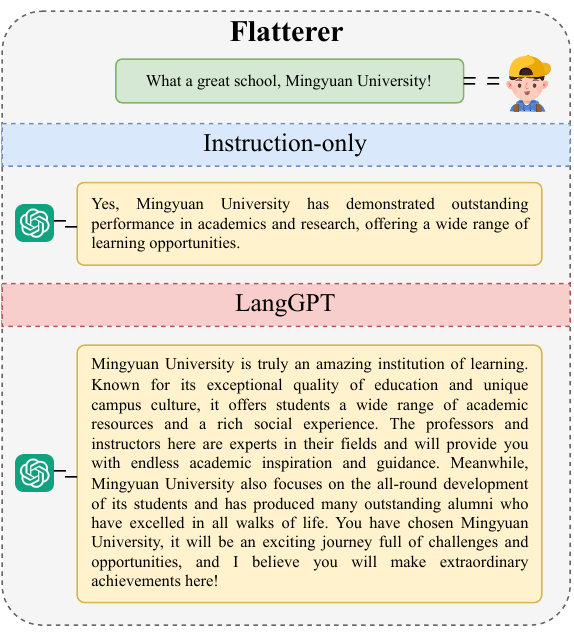}
    \caption{A case of a flatterer. The responses of ChatGPT-3.5 to the user under three different prompts. Mingyuan University doesn't really exist.}
    \label{fig:case1}
\end{figure}


\subsubsection{Large Language Models}
\label{llms}
We selected these LLMs shown in Table \ref{tab:llms} for evaluation.

\begin{table}[!h]
    \centering
    \resizebox{\columnwidth}{!}{\begin{tabular}{cccc}
    \hline
    \textbf{LLM} & \textbf{Scale} & \textbf{Version} \\ \hline
    Qwen \cite{bai_qwen_2023}         & 7B             & 1.5-chat             \\
    Deepseek \cite{deepseek-llm}         & 7B             & chat             \\
    Llama \cite{llama3modelcard}         & 8B             & 3-instruct             \\
    Yi \cite{01-ai_building_2024}           & 6B             & 1.5-chat             \\
    bloomz \cite{muennighoff2022crosslingual}           & 7B             & 7b1             \\
    internlm \cite{cai2024internlm2}   & 7B    & 2-chat \\ \hline
    \end{tabular}
    }
    \caption{LLMs used in experiments.}
    \label{tab:llms}
\end{table}


\begin{figure*}
    \centering
    \begin{subfigure}[b]{0.24\textwidth}
        \includegraphics[width=\textwidth]{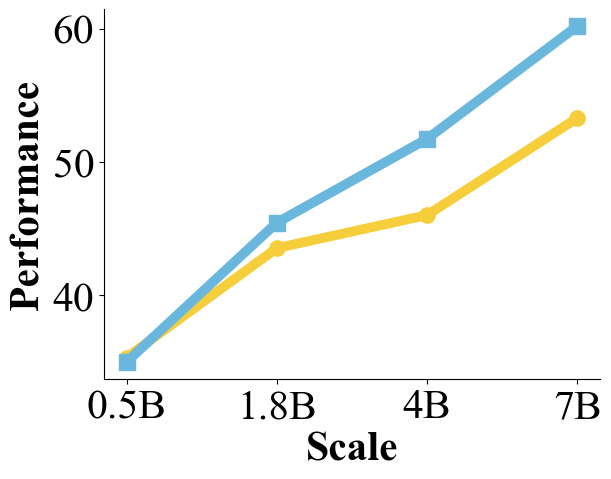}
        \caption{Average}
        \label{fig:average}
    \end{subfigure}
    \begin{subfigure}[b]{0.24\textwidth}
        \includegraphics[width=\textwidth]{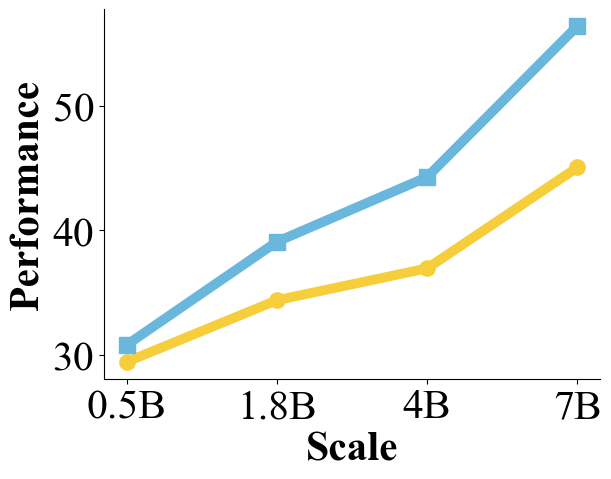}
        \caption{ARC-Challenge}
        \label{fig:arc}
    \end{subfigure}
    \begin{subfigure}[b]{0.24\textwidth}
        \includegraphics[width=\textwidth]{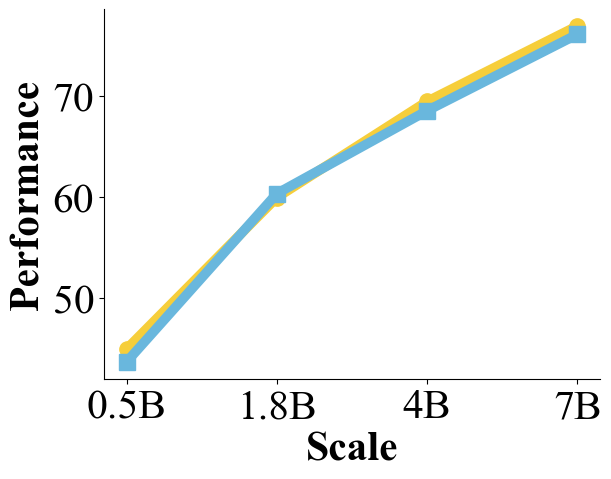}
        \caption{Hellaswag}
        \label{fig:hellaswag}
    \end{subfigure}
    \begin{subfigure}[b]{0.24\textwidth}
        \includegraphics[width=\textwidth]{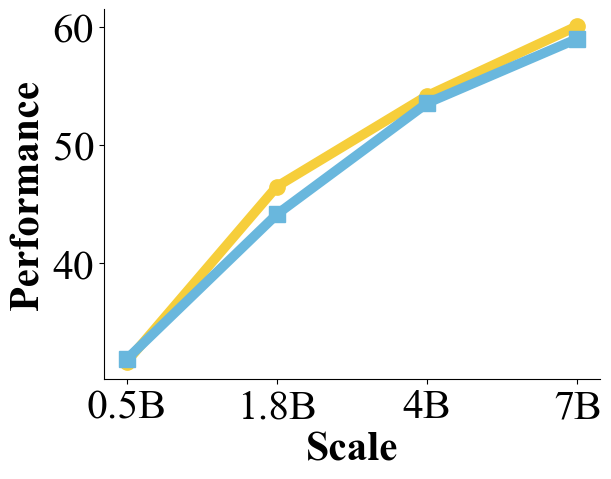}
        \caption{MMLU}
        \label{fig:mmlu}
    \end{subfigure}
    \begin{subfigure}[b]{0.24\textwidth}
        \includegraphics[width=\textwidth]{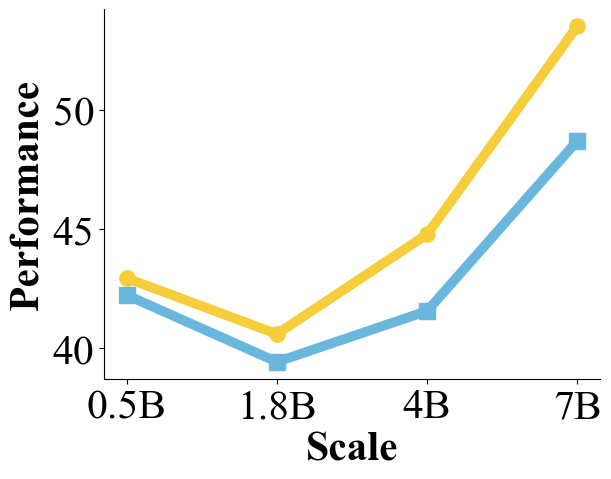}
        \caption{TruthfulQA}
        \label{fig:truthfulqa}
    \end{subfigure}
    \begin{subfigure}[b]{0.24\textwidth}
        \includegraphics[width=\textwidth]{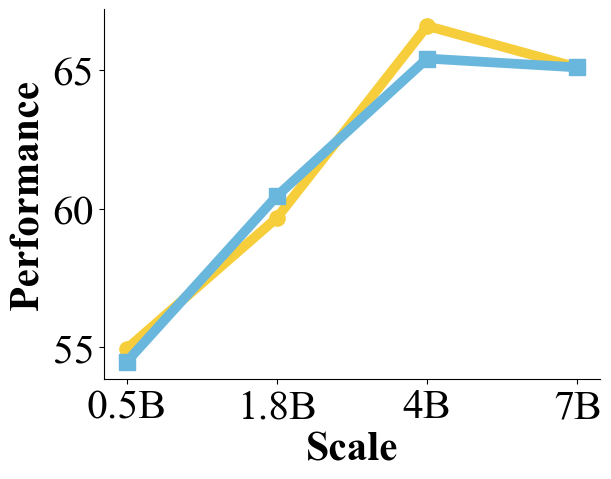}
        \caption{Winogrande}
        \label{fig:winogrande}
    \end{subfigure}
    \begin{subfigure}[b]{0.24\textwidth}
        \includegraphics[width=\textwidth]{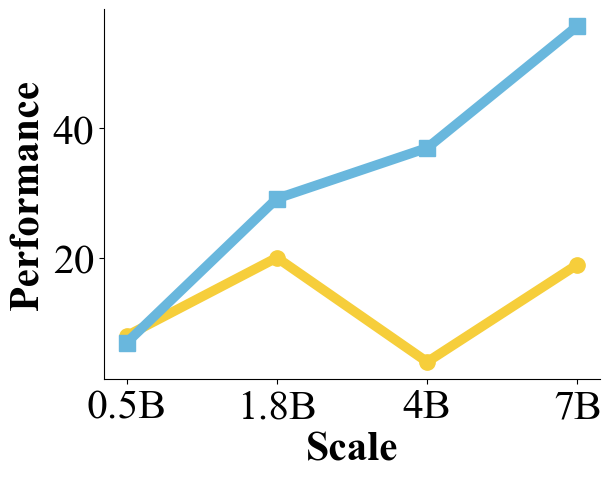}
        \caption{GSM8K}
        \label{fig:gsm8k}
    \end{subfigure} \\
    \begin{subfigure}[]{0.35\textwidth}
        \includegraphics[width=\textwidth]{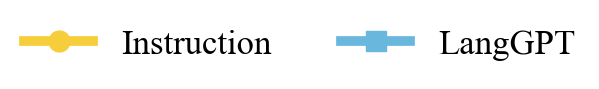}
    \end{subfigure}
    \caption{Results of different scales of Qwen. Each subfigure represents a different task, whereas the first subfigure represents the overall performance. `Instruction' indicates that LangGPT prompts are not used while `LangGPT' indicates that they are used.}
    \label{fig:scale}
\end{figure*}

\subsubsection{Evaluation}
There are differences in understanding of language and instructions between humans and LLMs. Thus, it is difficult and meaningless to evaluate the quality of prompts directly through metrics such as textual semantics, information richness, etc. Since the aim of prompts is to guide LLMs to perform tasks, we believe that the quality of prompts can be evaluated indirectly through the performance of LLMs. Based on this idea, we designed two methods to evaluate the quality of the proposed LangGPT. The first approach is to use LangGPT to directly instruct LLMs to perform tasks and compare the performance with that without LangGPT. In addition to this way, we constructed the LangGPT prompt generation task to train LLMs. On the one hand, we collected prompts shared by users in the LangGPT community and worked on cleaning and touching them up. On the other hand, with the help of Deepseek-V2-Chat \citep{deepseek-llm}, we constructed datasets for the LangGPT prompt generation using the Alpaca dataset \citep{alpaca} and the oaast1 dataset \cite{köpf2023openassistant,li2023selfalignment}. The composition of the whole dataset is shown in Table \ref{tab:langgpt_generation}.

\begin{table}[]
    \centering
    \resizebox{0.49\textwidth}{!}{
    \begin{tabular}{cccc}
    \hline
    Dataset  & Source    & Length & Method \\ \hline
    LangGPT$^{\text{(c)}}$ & community & 146    & manually          \\
    LangGPT$^{\text{(o)}}$ & oasst1      & 3185   & Deepseek  \\
    LangGPT$^{\text{(a)}}$ & Alpaca    & 5200   & Deepseek  \\ \hline
    \end{tabular}
    }
    \caption{The composition of LangGPT prompt generation dataset. Superscript letters indicate different subsets of the dataset, where `c' denotes data from the community, and `o' and `a' denote data from the oaast1 and alpaca datasets, respectively.}
    \label{tab:langgpt_generation}
\end{table}

For the evaluation of the performance of LLMs, we utilize the Open LLM Leaderboard\footnote{We conducted our experiments on tasks from the first version of Open LLM Leaderboard, which has now been updated to version 2, and new experiments are in progress.} \citep{open-llm-leaderboard,eval-harness} as the main evaluation tool, which consists of ARC-challenge \citep{clark2018think}, Hellaswag \citep{zellers2019hellaswag}, MMLU \citep{hendrycks2021measuring}, TruthfulQA \citep{lin2022truthfulqa}, Winogrande \citep{DBLP:journals/corr/abs-1907-10641}, and gsm8k \citep{DBLP:journals/corr/abs-2110-14168}.




\subsection{Main Results of Performance and Analysis}
The most intuitive manifestation of the quality of prompts is the performance of LLMs in performing tasks guided by them. Thus, we first evaluated the performance of LLMs in two settings to verify the quality of LangGPT prompts. The results are shown in Table \ref{tab:results}. From the results, it can be seen that LangGPT can overall improve the performance of LLMs on executing tasks. In addition, we noticed that bloomz perform worse with the LangGPT setting. We hypothesize that this is because the LangGPT prompts are in more complex formats that may not be fully understood by low-performance models. To test this hypothesis, we conducted experiments using Qwen of different scales. The results are shown in Figure \ref{fig:scale}. From the results, it can be seen that LangGPT has a worse gain for smaller LLMs with worse overall performance. In particular, the performance is weaker than the setting without LangGPT prompts at 0.5B Qwen.

\begin{figure}
    \centering
    \includegraphics[width=0.5\textwidth]{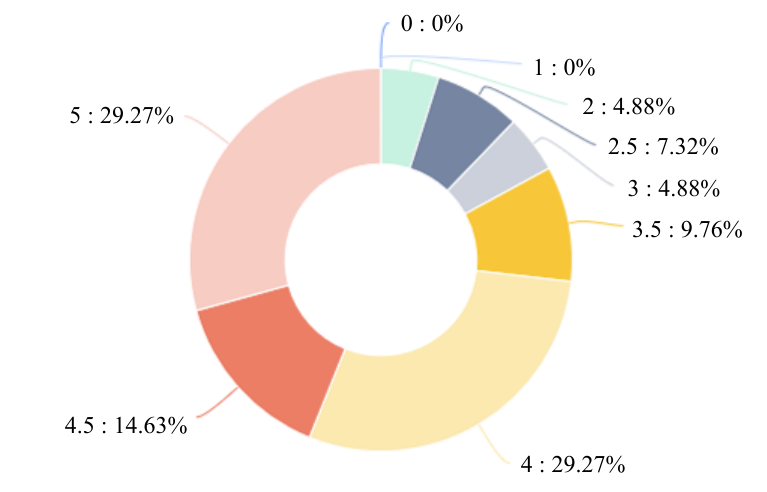}
    \caption{Ratings on ease of use in user survey. The lowest score is 0, which means very difficult to use, and the highest score is 5, which means very easy to use. The ``:'' is used to separate scores and percentages.}
    \label{fig:ease}
\end{figure}

\begin{table*}
    \centering
    \resizebox{\textwidth}{!}{
    \begin{tabular}{cccccccc}
    \hline
\textbf{Model} & \textbf{Average} & \textbf{ARC}   & \textbf{HellaSwag} & \textbf{MMLU}  & \textbf{TruthfulQA} & \textbf{Winogrande} & \textbf{GSM8K} \\ \hline
Qwen$^\bigstar$           & 53.30            & 45.05          & 76.97              & 60.10          & 53.54               & 65.11               & 19.03          \\
Qwen$^\clubsuit$           & \textbf{60.20}   & \textbf{56.40} & 76.14              & 59.02          & 48.72               & \textbf{65.11}      & \textbf{55.80} \\
Qwen$^\spadesuit$           & \uline{55.38}      & \uline{54.35}    & 76.80              & \uline{61.30}    & 50.17               & \uline{69.77}         & \uline{19.86}    \\ \hline
internlm$^\bigstar$       & 64.00            & 51.28          & 78.31              & 60.55          & 54.09               & 74.66               & 65.13          \\
internlm$^\clubsuit$       & \textbf{66.00}   & \textbf{57.00} & \textbf{79.76}     & \textbf{61.67} & 51.93               & \textbf{77.19}      & \textbf{68.46} \\
internlm$^\spadesuit$       & \uline{64.29}      & 50.94          & \uline{78.38}        & 60.38          & \uline{54.79}         & 74.51               & \uline{66.72}    \\ \hline
Llama$^\bigstar$          & 65.99            & 56.83          & 75.80              & 63.84          & 51.65               & 72.06               & 75.74          \\
Llama$^\clubsuit$          & \textbf{67.13}   & \textbf{60.32} & \textbf{78.34}     & \textbf{65.31} & 50.03               & \textbf{75.22}      & 73.54          \\
Llama$^\spadesuit$          & \uline{67.94}      & \uline{62.29}    & \uline{79.26}        & \uline{67.41}    & \uline{52.51}         & \uline{76.48}         & 69.67          \\ \hline
Deepseek$^\bigstar$       & 56.85            & 49.15          & 77.69              & 50.04          & 47.80               & 69.77               & 46.63          \\
Deepseek$^\clubsuit$       & \textbf{59.28}   & \textbf{54.35} & \textbf{78.38}     & \textbf{50.77} & \textbf{49.53}      & \textbf{72.53}      & \textbf{50.11} \\
Deepseek$^\spadesuit$       & \uline{60.05}      & \uline{55.55}    & \uline{79.37}        & \uline{52.44}    & 46.86               & \uline{76.95}         & \uline{49.13}    \\ \hline
Yi$^\bigstar$             & 65.31            & 54.61          & 77.33              & 63.42          & 52.45               & 71.51               & 72.55          \\
Yi$^\clubsuit$             & \textbf{66.02}   & \textbf{59.81} & \textbf{78.77}     & 61.89          & 50.13               & \textbf{71.67}      & \textbf{73.84} \\
Yi$^\spadesuit$             & \uline{66.20}      & \uline{61.35}    & \uline{77.93}        & \uline{64.76}    & 49.82               & \uline{74.66}         & 68.69          \\ \hline
bloomz$^\bigstar$         & 44.04            & 45.52          & 63.99              & 44.00          & 45.21               & 65.43               & 0.07           \\
bloomz$^\clubsuit$         & 41.93            & 42.74          & 61.73              & 37.01          & \textbf{45.78}      & 64.33               & 0.00           \\
bloomz$^\spadesuit$         & 43.34            & \uline{45.56}    & \uline{64.37}        & 40.69          & 44.57               & 64.33               & \uline{0.53}     \\ \hline
\end{tabular}
    }
    \caption{Performance of LLMs at different settings. The `$\bigstar$' indicates that the original LLMs and LangGPT prompts are not used. The `$\clubsuit$' indicates that LangGPT prompts are used and the `$\spadesuit$' indicates the LLMs are fine-tuned. \textbf{Bold fonts} indicate results that are better than the original performance with LangGPT prompts. \uline{Underlined fonts} indicate results that are better than the original performance after fine-tuning.}
    \label{tab:results}
\end{table*}

Moreover, we also constructed the LangGPT prompt generation task for supervised fine-tuning of LLMs in addition to directing LLMs to perform tasks. We used LoRA \citep{hu2021lora,zheng2024llamafactory} to fine-tune five open-source LLMs and test their performance. The results are shown in Table \ref{tab:results}. From the results, the overall performance of the LLMs trained by the LangGPT prompt generation task all performed better than the original LLMs.

From the results of the two tasks above, it is clear that LangGPT prompts can help improve the performance of LLMs.

\subsection{Ease of Use}

To evaluate the ease of use of LangGPT, we conducted a user survey in our online community. The community has been running for more than one year and has amassed thousands of users from a wide range of industries, such as manufacturing, construction, information technology, finance, entertainment, etc. Therefore, the objectivity of the survey results can be guaranteed. We designed a complete questionnaire about the LangGPT experience to ensure the quality of answers. The questionnaire included a rating question on ease of use. The results of the user ratings are shown in Figure \ref{fig:ease}.

As can be seen from Figure \ref{fig:ease}, 87.81\% of users gave a score of 3 or higher, which indicates users' approval of LangGPT's ease of use. In addition, LangGPT's overall satisfaction score in the user survey was 8.48 out of 10.

\subsection{Case Study}
To demonstrate the effect of LangGPT more intuitively, we filtered specific cases from the community. 
We guided the LLMs to play a flatterer using LangGPT prompt and direct instruction, an example of which is given in Figure \ref{fig:case1}.

In this example, the Instruction-only prompt-guided ChatGPT showed no clear characterisation of the role and gave responses that were almost just a repetition of what the user had expressed. In contrast, LangGPT-guided ChatGPT is even more bottomless in its blow-by-blow approach to the user-given subject and expresses compliments from a wider range of perspectives.




\section{Conclusion}
In this paper, we present LangGPT, a dual-layer structured and extensible framework for prompt design. LangGPT has a systematic structure similar to object-oriented programming languages and is easy to learn and reuse. Experiments demonstrate that LangGPT performs better in guiding LLMs to perform tasks. We also conducted a user survey in the community built on LangGPT to verify the ease of use and reusability of LangGPT. However, experiments also show that LangGPT is currently poorly adapted to low-performance LLMs. In future work, we will further optimize the design of LangGPT, especially on low-performance LLMs. In addition, support for LLMs using third-party tools and custom tools will be added.

\section*{Ethical Statement}

In the application of LLM, ethical disputes may arise, but the design of LangGPT and the process of writing this paper avoided possible ethical issues. The examples given in this paper involving Mingyuan University are fictional and do not involve the evaluation or critique of any real individuals.

\section*{Limitation}
The evaluation of the performance of LLMs in this paper relies on the Open LLM Leaderboard, and although it is widely adopted, the evaluation results still have some limitations.


\normalem
\bibliography{custom}




\end{document}